\documentclass[english,a4paper,aps,prl,twocolumn,groupedaddress]{revtex4}
\usepackage[T1]{fontenc}
\usepackage[latin1]{inputenc}
\usepackage{amsmath}
\usepackage{graphicx}
\usepackage{amssymb}
\usepackage{color}



\usepackage{babel}
\usepackage{color}
\makeatother
\begin{document}

\title{Nonlinear Dynamics of Magnetic Islands Imbedded in Small Scale
  Turbulence}

\author{M. Muraglia$^{1}$, O. Agullo$^{1}$, S. Benkadda$^{1}$,
  X.  Garbet$^{2}$, P. Beyer$^{1}$, A. Sen$^{3}$ }

\affiliation{$^{1}$France-Japan Magnetic Fusion Laboratory, LIA 336 CNRS, Marseille, France\\
$^{2}$CEA, IRFM, 13108, St-Paul-Lez-Durance, France\\
$^{3}$Institute for Plasma Research, Bhat, Gandhinagar 382428, India}

%

\begin{abstract}
The nonlinear dynamics of magnetic tearing islands imbedded in a
pressure gradient driven turbulence is investigated numerically 
in a reduced magnetohydrodynamic model. The study reveals regimes
where the linear and nonlinear phases of the tearing instability are
controlled by the properties of the pressure gradient.
In these regimes, the interplay between the pressure and the magnetic 
flux determines the dynamics of the saturated state. A secondary instability 
can occur and strongly modify the magnetic island dynamics by triggering a
poloidal rotation. It is shown that the complex nonlinear interaction
between the islands and turbulence is nonlocal and involves small
scales.
\end{abstract}
\maketitle Magnetic reconnection is a complex phenomenon involving plasma flows and a rearrangement of the magnetic field lines inside a narrow region (the reconnection layer) where topologically different magnetic flux tubes can get interconnected and reconfigure themselves. It plays an important role in fusion experiments and in many astrophysical events \cite{Biskamp0093}. In a complex fusion device, such as a tokamak, the plasma is susceptible
to many kinds of instabilities which can occur concurrently at various 
space and time scales. Such a coexistence of microturbulence and magnetohydrodynamic
(MHD) activities has been observed in many experiments \cite{Tanaka05} 
with some evidence of correlated effects arising from their simultaneous existence.
An important question to address is therefore the nature and amount of mutual interaction
between microturbulence and large-scale MHD instabilities - an issue that is at the heart
of multi-scale phenomena of complex systems in astrophysics,
geophysics, nonlinear dynamics and fluid turbulence. Early analytic attempts at investigation of this important question have relied on ad-hoc modeling of turbulence effects through anomalous transport coefficients \cite{Kaw79}. More recently a minimal self-consistent model based on wave kinetics and adiabatic theory has been used in \cite{Devitt06} to study the interaction of a tearing mode with drift wave turbulence. 
Numerical simulation studies in \cite{Baar05} have directly addressed the problem of multiscale interactions and have taken into account the nonlinear modifications of the equilibrium profiles due to turbulence. Such studies have been extended in \cite{Ishizawa07} to investigate the interaction between double tearing modes and micro-turbulence through the excitation of zonal flows.
Finally in \cite{Militello08} a numerical investigation of the interaction of a 2D electrostatic turbulence with an island whose dynamics is not fully self-consistent but is governed by a generalized Rutherford equation has been carried out. In this paper we report on self-consistent simulations of the multiscale interaction between microturbulence driven by pressure gradients and magnetic islands with a focus on regimes where the growth of the latter is essentially due to pressure effects and where small-scale dynamics appear to be important.  
The background microturbulence is found to induce a nonlinear rotation of the island as well as to significantly alter its final quasi-equilibrium state by the excitation of a secondary instability. We discuss the characteristics of the various stages of the nonlinear evolution and also delineate the role of small scales in the overall dynamics of the system.

{
 We consider a minimalist two-dimensional plasma model based on the two fluid Braginskii equations in the drift approximation \cite{Scott85,Ottaviani04} with cold ions and isothermal electrons. The model includes magnetic curvature effects and electron diamagnetic effects but neglects electron inertia and Hall effect contributions.
The evolution equations are }
\begin{equation}
  \frac{\partial}{\partial
    t}\nabla_{\perp}^{2}\phi+\left[\phi,\nabla_{\perp}^{2}\phi\right]=\left[\psi,\nabla_{\perp}^{2}\psi\right]-\kappa_{1}\frac{\partial
    p}{\partial
    y}+\mu\nabla_{\perp}^{4}\phi,\label{eq:equPHI}\end{equation}

\begin{multline}
\frac{\partial}{\partial
  t}p+\left[\phi,p\right]=-v_{\star}\biggl((1-\kappa_{2})\frac{\partial\phi}{\partial
  y}+\kappa_{2}\frac{\partial p}{\partial
  y}\biggr)\\ +\hat{\rho}^{2}\left[\psi,\nabla_{\perp}^{2}\psi\right]+\chi_{\perp}\nabla_{\perp}^{2}p,\label{eq:equPE}\end{multline}

\begin{equation}
\frac{\partial}{\partial
  t}\psi=\left[\psi,\phi-p\right]-v_{\star}\frac{\partial\psi}{\partial
  y}+\eta\nabla_{\perp}^{2}\psi,\label{eq:equPSI}\end{equation}

\noindent where the dynamical field quantities are the electrostatic potential $\phi$, the electron pressure $p$ and the magnetic 
flux $\psi$. 
The equilibrium quantities are a constant pressure gradient and a magnetic field corresponding to a Harris current sheet model \cite{Biskamp0093}. Further, 
$\kappa_{1}=2\Omega_{i}\tau_{A}\frac{L_{\perp}}{R_{0}}$ and
$\kappa_{2}=\frac{10}{3}\frac{L_{p}}{R_{0}}$ are the curvature terms
with $R_{0}$ representing the major radius of a toroidal plasma configuration.
$L_{p}$ is the gradient scale length,
$\tau_{A}$ is the Alfv\'en time based on a reference perpendicular
length scale $L_{\perp}$ and $\Omega_{i}$ is the ion cyclotron
frequency. Equations (\ref{eq:equPHI}-\ref{eq:equPSI}) are normalized
using the characteristic Alfv\'en speed $v_{A}$ and the length scale
$L_{\perp}$. 
$\mu$ is the viscosity, $\chi_{\perp}$ the perpendicular
diffusivity, $\eta$ is the plasma resistivity,
$v_{\star}=\beta_{e}/\Omega_{i}\tau_{A}$ is the normalized electron
diamagnetic drift velocity with $\beta_{e}$ being the ratio between
the electronic kinetic pressure and the magnetic pressure.
 $\hat{\rho}=\frac{\rho_{S}}{L_{\perp}}$ is the normalized ion sound Larmor radius.
In the limit
$R_{0}\rightarrow \infty$, we recover the drift tearing model
\cite{Ottaviani04}, and when magnetic fluctuations are weak,
$\psi\sim0$ $(\kappa_{1}\ne0$), the system describes the electrostatic
interchange instability. Conversely, the large island limit,
$\hat{\rho}=v_{\star}=0$ with $\Omega_{i}\tau_{A}\sim1$, gives the high $\beta$
model which was originally introduced by H. R. Strauss
\cite{Strauss77}.  
The minimalist model used here is, in fact, a reduced version of the
four fields model of \cite{Hazeltine85_76} where we have ignored the parallel ion dynamics and thereby neglected its effect on the transversal pressure balance \cite{Hazeltine85_76}.

As a preliminary to the numerical study of eqs.(\ref{eq:equPHI}-\ref{eq:equPSI}), we first look at some linear results of a 
simplified set of equations with $\mu=\chi_{\perp}=\kappa_{i}=0$ where
it is possible to obtain analytic relations for the linear growth rate of the tearing
mode under the constant $\psi$ approximation \cite{Biskamp0093,White01}, namely, 

\begin{equation}
\begin{array}{ccc}
  \Delta' & = &
  \frac{\gamma^{2}}{k_{y}^{2}}\alpha^{-3}\int_{-\infty}^{+\infty}\frac{\chi''\left(z\right)}{z}dz,\\ z
  & = &
  -z^{2}\chi\left(z\right)+(1+\hat{\rho}^{2}\alpha^{-2}z^{2})\chi^{''}\left(z\right)\end{array}
\label{eq:DeltaPrime}
\end{equation}
where
$\phi\left(x\right)=-\alpha^{-1}\psi\left(0\right)\chi\left(z\right)$,
$z=\alpha x$,
$\alpha=\left(\eta\gamma/k_{y}^{2}\right)^{1/4}$ and $\Delta^{\prime}$ is the standard
stability parameter. The perturbed pressure is given by
$p\left(x\right)=\hat{\rho}^{2}\phi''\left(x\right)$ (with $\hat{\rho}=0$
corresponding to the classical tearing situation).
\begin{figure}
\hskip-0.75cm
\includegraphics[width=9.cm,height=4.5cm]{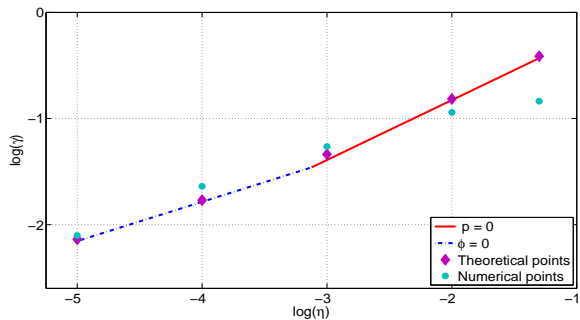}
\vskip.0cm
\caption{Numerical and
 theoretical results of the linear growth rate $\gamma$ versus $\eta$ at
 $\nu=\chi_{\perp}=0$, $\Delta'=6$ and $\hat{\rho}=10^{-1}$. \label{cap:figure1}}
\end{figure}
Fig.~\ref{cap:figure1} shows the dependence of the linear growth
rate $\gamma(\eta)$ of the instability on the resistivity with
$\hat{\rho}=10^{-1}$. The numerical results (circles) are seen to agree quite well
with the values (diamonds) of the solution to the analytic relation 
Eq.(\ref{eq:DeltaPrime}). We observe that $\gamma(\eta)$ exhibits a change of slope
after a certain value of $\eta$. The two regimes correspond to the two 
limiting cases $T_{e}=0$ or $p=0$ (solid
line) and $\phi=0$ (dashdot line). The intersection of the two lines
gives the critical value of the resistivity
$\eta_{c}=0.58\;\Delta'^{-1/2}\hat{\rho}^{5/2}\sim8\times 10^{-4}$ for $\Delta^{\prime} = 6$. When
$\eta>\eta_{c}$, the linear growth rate given by the classical tearing
case is higher than the other limiting case. 
The system chooses the more unstable case and
the classical tearing mode is recovered with the scaling laws
$\gamma\sim\eta^{3/5}$ and $\delta\sim\eta^{2/5}$. When
$\eta<\eta_{c}$, the coupling between $p$ and $\psi$ is strong and the
island formation is driven by the pressure perturbation. The resistive
layer becomes thinner or more singular than in the case $T_{e}=0$
where only a $(\phi,\psi)-$coupling exists.  
Further the disagreement
observed between the numerical and the theoretical results for
$\eta>5\times 10^{-2}$ is a consequence of the breaking down of the constant
$\psi$ approximation in this regime.

 
We now discuss the full nonlinear numerical simulation of 
Eqs.(\ref{eq:equPHI}-\ref{eq:equPSI}) that explores the mutual 
interaction between small scale interchange modes and a small magnetic
island. A semi-spectral code with a 2/3-dealiasing rule in the poloidal
direction, a resolution of
$128$ grid points in the radial direction, $96$ poloidal
modes and that maintains conservation properties of the nonlinear
terms to a high degree, has been used. The computational box size is $L_{x}=L_{y}=2\pi$.  In order to isolate the
nonlinear mechanisms responsible for the island rotation, the linear
diamagnetic effect has been turned off in Eq.(\ref{eq:equPSI}). The effect of the latter on the evolution of the tearing mode is well known, namely that it leads to a real frequency and consequently a rotation of the island in the diamagnetic drift direction. Note that we have checked \textit{\`a posteriori} by turning on the linear diamagnetic term in eq.(3) that the amount of induced nonlinear rotation (obtained by subtracting the linear diamagnetic frequency from the total rotation) remains the same. In eq.(3) we also set
$\kappa_{2}=0$, since we find from our simulations 
that the $\kappa_{2}$ contribution is rather weak.  $\hat{\rho}$ and
$v_{\star}$ are taken to be equal to $1$ and $\beta_{e}=10^{-2}$.  The
parameter related to the interchange instability is
$\kappa_{1}=10^{-2}$.  The shape of the equilibrium magnetic field is 
chosen to allow a tearing instability 
to develop with a poloidal mode number $k_{y}=1$ with $\Delta'=6$\cite{Biskamp0093}. 
\begin{figure}
\hskip-0.75cm
\includegraphics[width=9.cm,height=4.5cm]{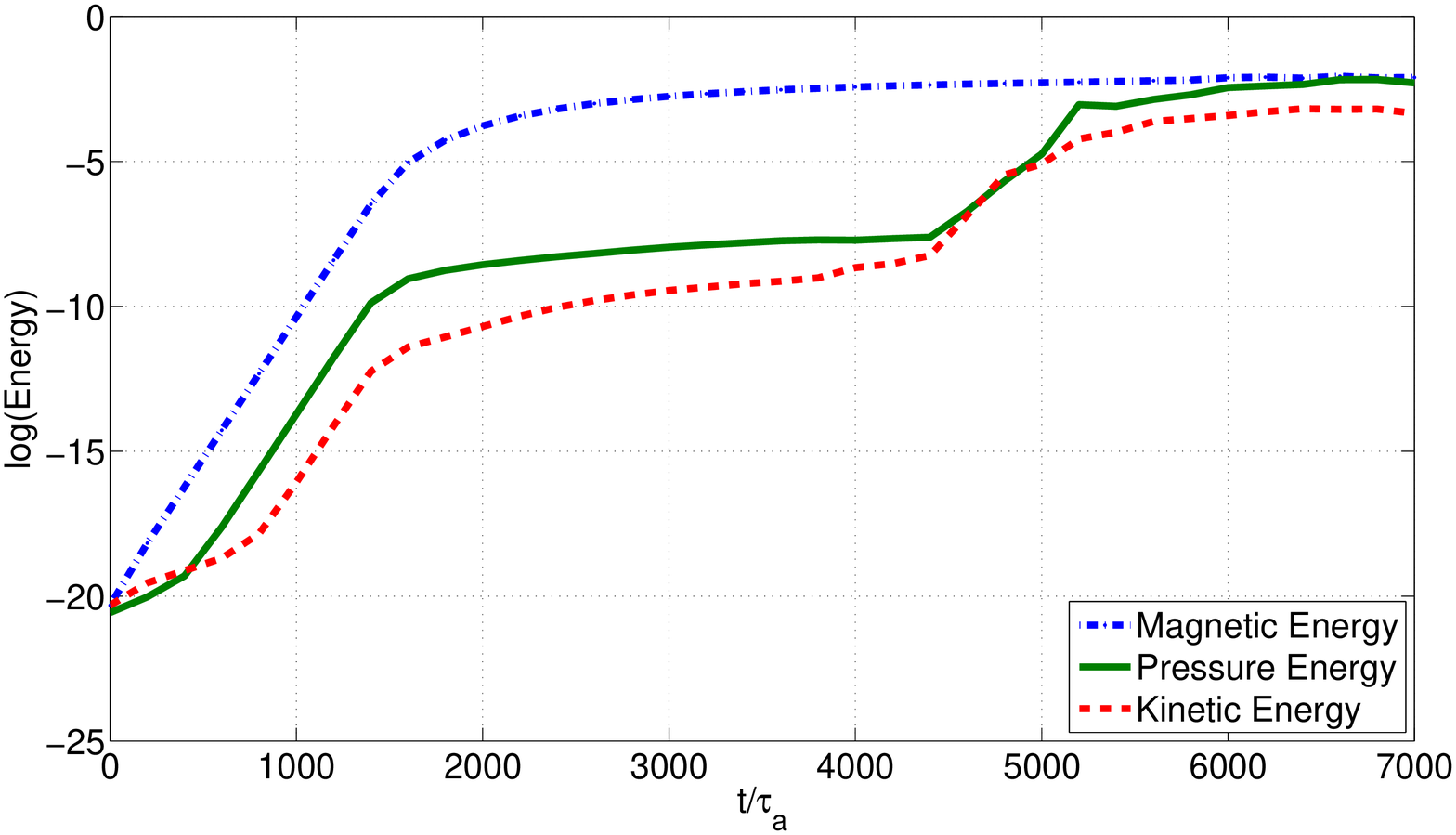}
\vskip.0cm
\caption{Time evolution of the magnetic, pressure and kinetic energies.\label{cap:figure2}}
\end{figure}
Fig.~\ref{cap:figure2} shows, for $\mu=\chi_{\perp}=\eta=10^{-4}$,
the time evolution of the magnetic ($E_\psi$), pressure ($E_p$) and kinetic ($E_\phi$) energies of the fluctuations for $\eta<\eta_{c}$,
corresponding to a regime where the magnetic island generation is
pressure driven. 
Four phases are observed. First, a exponential growth of
the magnetic island ($t\lesssim1300\tau_{A}$), followed by a
quasi-plateau phase with however an increase of the energies of the
three fields ($t\lesssim4500\tau_{A}$).  Next, a phase characterized
by an abrupt growth of the kinetic and pressure energies in
which the kinetic energy level equals the energy of pressure
perturbations and finally, the system reaches a new quasi-plateau phase
for ($t\gtrsim5100\tau_{A}$). 
During the linear and first plateau phases, the energy
associated with the pressure perturbations is higher than the kinetic
energy, \emph{i.e} the dynamics is controlled by an interplay between
the magnetic flux and the pressure. In the second phase,
$t\lesssim4500\tau_{A}$, the magnetic island is maintained by adjacent
pressure cells similar to what is usually observed for flow cells in
the nonlinear regime of a tearing island \cite{Biskamp0093}.
This is illustrated in Fig.~\ref{cap:figure34} (upper panel,
$t=3000\tau_{A}$). During this phase, the kinetic energy piles up in
the flow cells which are located in the vicinity of the island. 
After $t\sim 3600\tau_A$, the flow cells are
no longer located in the vicinity of the magnetic island.
At $t\gtrsim4500\tau_{A}$,  a sharp growth of the kinetic and
pressure energies occurs. 
Far from the island the current is not significant, and for $t/\tau_{A}\in[4500,5000]$, a dominant interchange mode outside the sheet $(\phi_{11}, p_{11})$  is enhanced. The associated kinetic and pressure energies of the latter are equal. Here, $\phi_{11}$ means $\phi(k_x=1,k_y=1)$. The competition between the interchange and tearing modes  lead to the generation of small scale pressure structures in the vicinity of the island that
suffer further destabilization leading
to a drastic modification of the dynamics. Indeed, in less than $200$ Alfven times,
around $t\sim5000\tau_{A}$, an abrupt growth of the energy contained
in the pressure perturbation is observed and the system dynamics changes,
\emph{i.e}, a bifurcation occurs.
\begin{figure}
\includegraphics[scale=0.25]{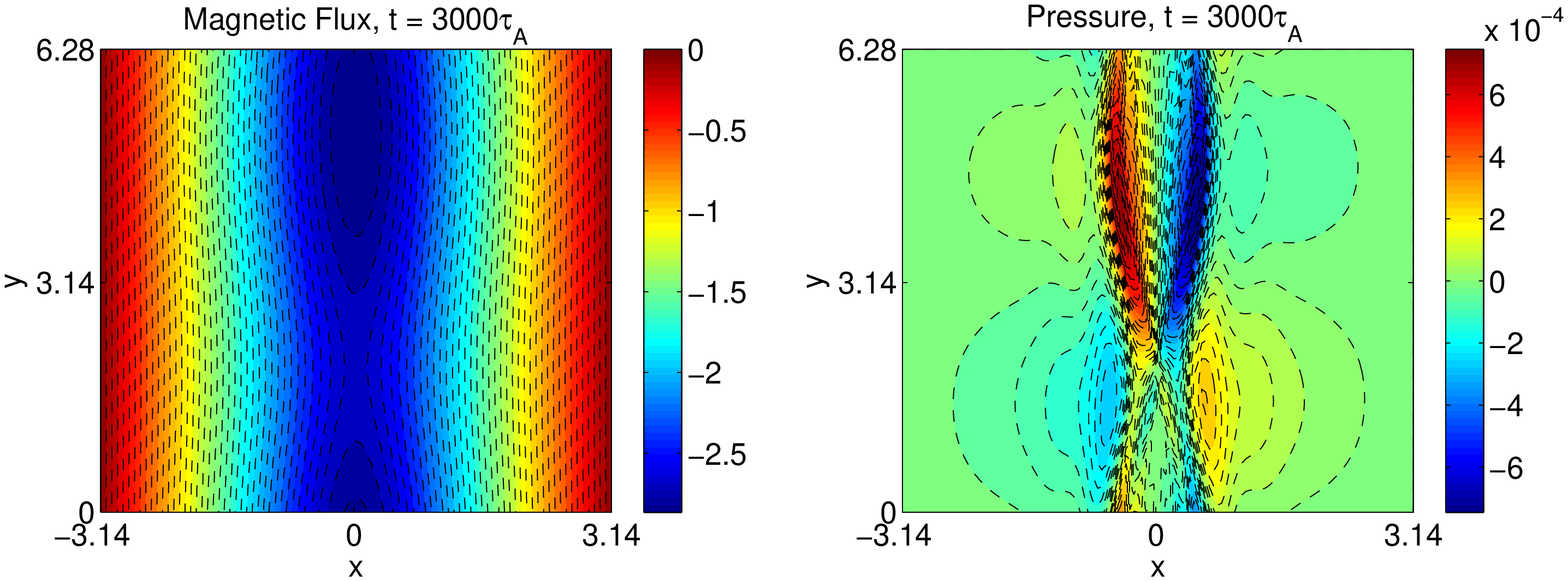}
\includegraphics[scale=0.25]{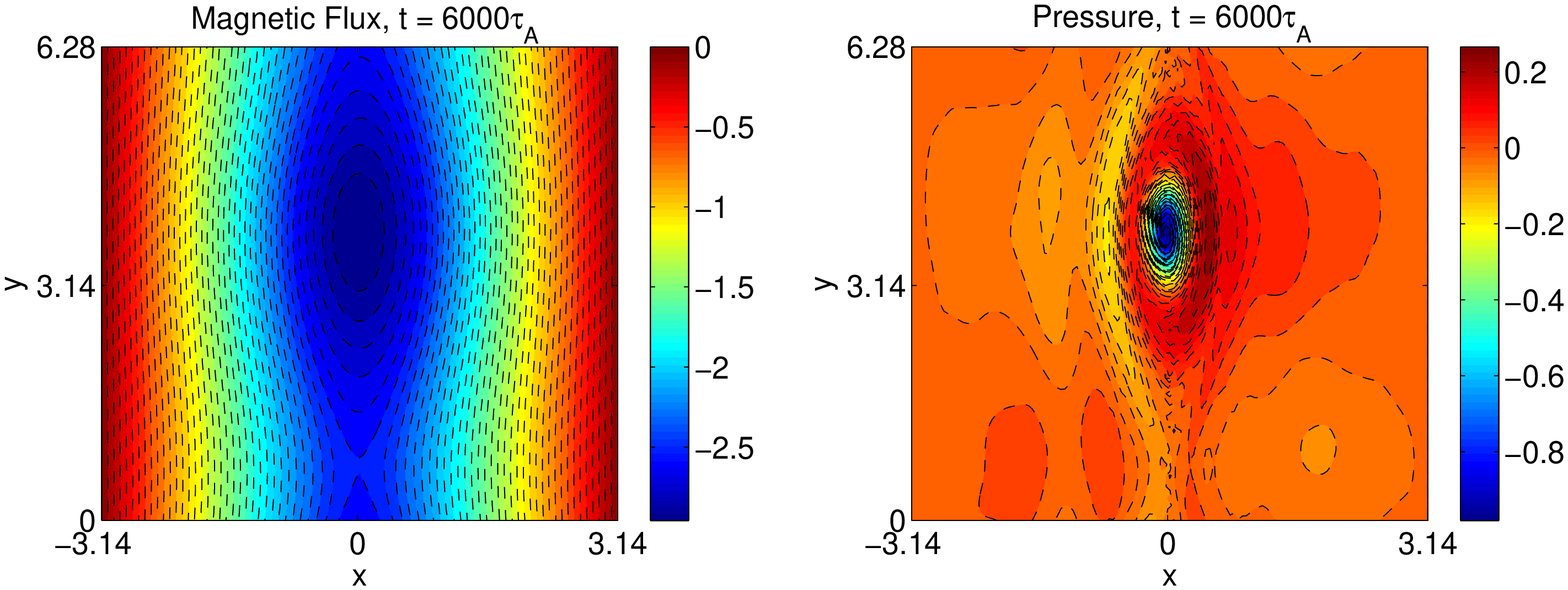}
\caption{Snapshots of the magnetic flux and the pressure at
  $t=3000\tau_{A}$ (Upper panel) and $t=6000\tau_{A}$ (Lower
  panel). \label{cap:figure34}}
\end{figure}
At larger times, $t>5100\tau_{A}$, the pressure dominates over the
flow, $E_{p}\gg E_{\phi}$, and the size of the magnetic island finally
saturates. Fig.~\ref{cap:figure5} shows the energy spectra of the
fields just before 
and after the bifurcation
. Before the bifurcation, the interchange mode is observed at $k_{y}=1$,
and as long as $k_{y}\ge2$,
the pressure energy is much higher than the kinetic energy. 
After this dynamical
bifurcation, we observe a persistence of small scales as well as
an enhancement of the energies (Fig.~\ref{cap:figure5}b). We also
find that a mean poloidal pressure and flow have been generated, and
as we will see below, it is linked to the rotation properties of the
island. For $8\le k_{y}\le50$, there is a trend towards an
equipartition of the magnetic and pressure spectra. It is worth noting
that even though the magnetic island is still, at this point, in a quasilinear
stage (the magnetic energy being concentrated on the mode
$k_{y}=1$), the pressure perturbation has a fully nonlinear structure and is made up
mainly of the modes $k_{y}<7$. 
\begin{figure}
\hskip-.75cm
\includegraphics[width=9.cm,height=4.cm]{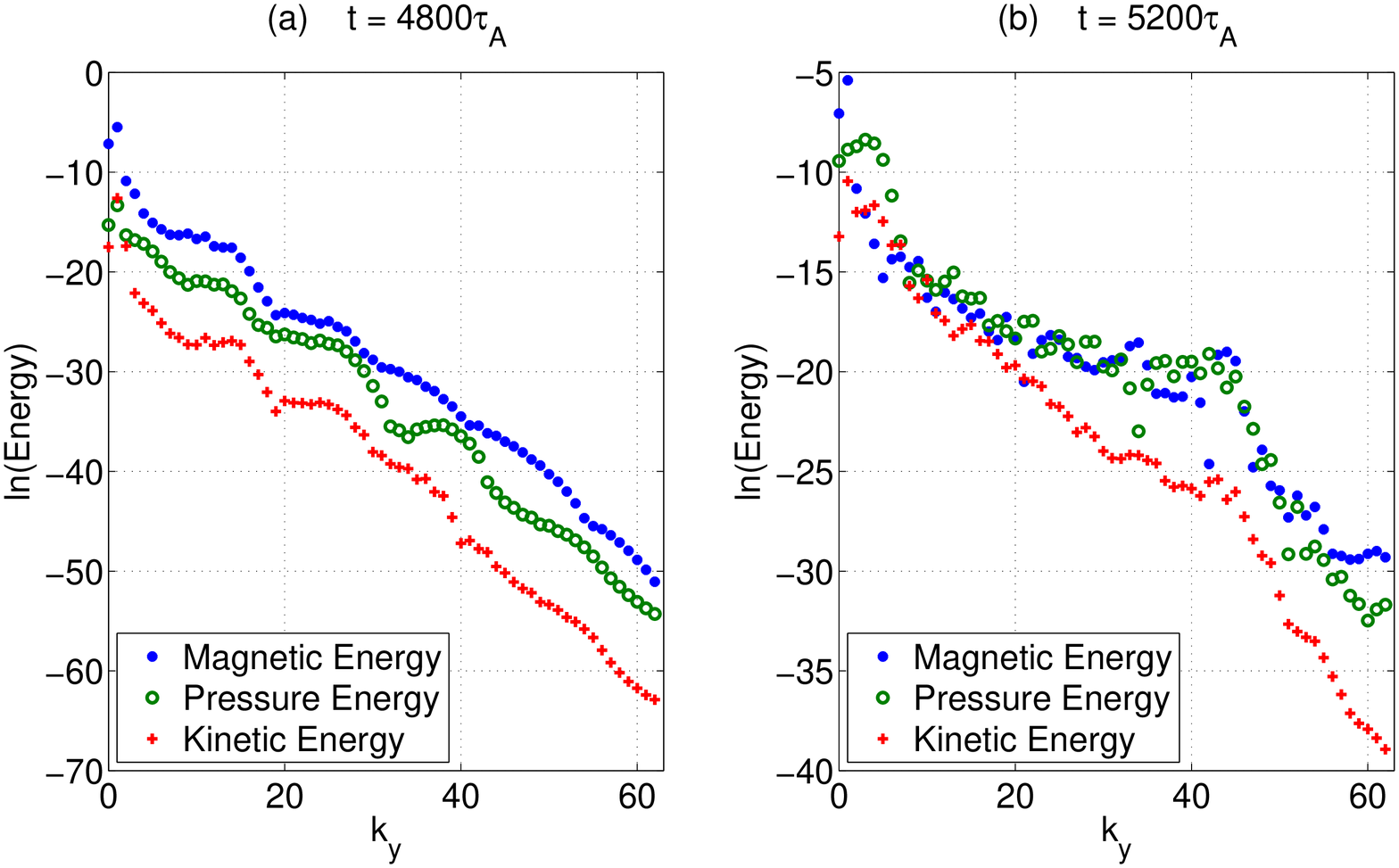}
\caption{Spectral energy densities as functions of the poloidal mode
 number $k_{y}$, just (a) before 
 ( at $t=4800\tau_{A}$) and (b) after (at
 $t=5200\tau_{A}$), the bifurcation. \label{cap:figure5}}
\end{figure}
An interesting feature clearly observed in the snapshot shown in Fig.~\ref{cap:figure34} at $t=6000\tau_{A}$ 
 is the generation of an island structure in the pressure field 
containing almost $90\%$ of the pressure energy $E_p$.

\begin{figure}
\hskip-0.75cm
\includegraphics[width=9.0cm,height=4.cm]{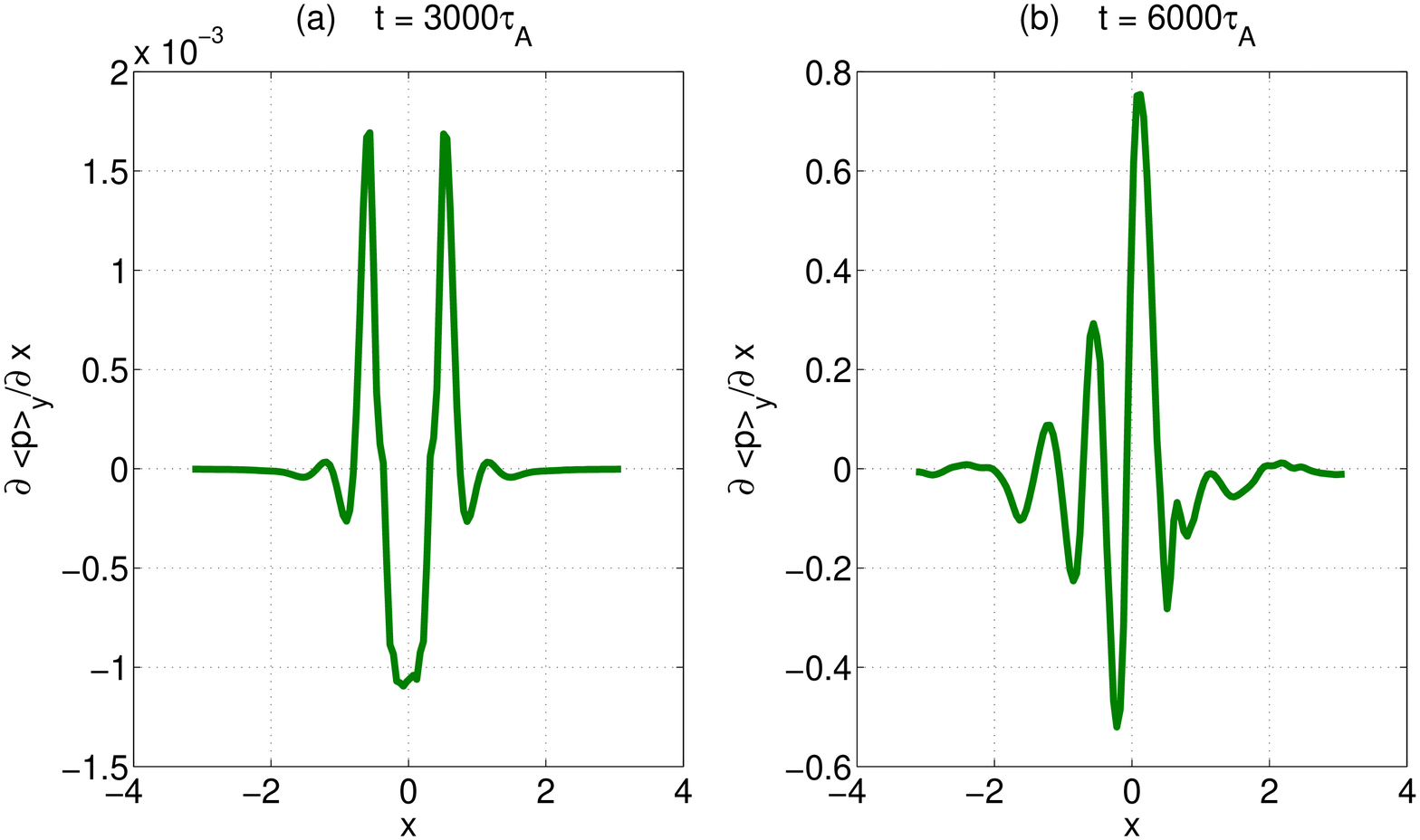}
\vskip.0cm
\caption{Plots of the poloidal diamagnetic velocity $v_{dia}$ at
  $t=3000\tau_{A}$ (a) and $t=6000\tau_{A}$
  (b).\label{cap:figure6}}
\end{figure}
\begin{figure}
\includegraphics[width=9cm,height=4.5cm]{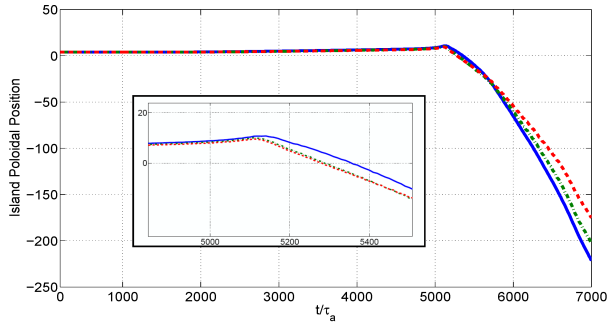}
\caption{Time evolution of the poloidal position of the center of the
  magnetic island $y_{island}(t)$ (solid blue line), and the models
  $y_{dia}(t)$ (dashed red line), $y_{p,\phi}(t)$
  (dashdot green line).\label{cap:figure7}}
\end{figure}

In the final stage where the energies reach a new quasi-plateau, as observed in Fig.~\ref{cap:figure2},
the change of dynamics is characterized by two important macroscopic features. First, there is a change of symmetry - the
poloidal diamagnetic velocity $v_{\mbox{dia}}=\frac{\partial}{\partial x}<p>_{y}$ having even parity for $t\lesssim5100\tau_{A}$ (brackets mean an average over the poloidal direction), loses this property after the bifurcation and has in fact an odd parity in the vicinity of the current sheet. This change of parity is clearly shown in Fig.~\ref{cap:figure6}) where  $v_{dia}$ is plotted, before and after the transition, at $t=3000\tau_{A}$ and $t=6000\tau_{A}$ respectively. The second macroscopic change is the inversion of the poloidal
rotation direction of the magnetic island together with an amplification of the velocity. 
{
The amplified velocity arising from the nonlinear interactions is of the order of the linear diamagnetic velocity as verified from \textit{\'a posteriori} runs made with the linear diamagnetic term retained in the equations. The change of direction in the island rotation can be observed in the zoomed frame of 
Fig.~\ref{cap:figure7} which shows the time evolution of the poloidal position of the island.
We find that the increase of $v_{dia}$ at the transition is linked to the coincident growth of the interchange mode ($\phi_{11}$,$p_{11}$) which feeds the angular momentum. The detailed mechanism of this nonlinear generation of angular momentum is however not known at this time and remains an open question.}


Some insights into the origin of the island poloidal rotation can be
obtained from Eq.(\ref{eq:equPSI}) where one notes that 
both the self generated zonal and diamagnetic flow terms, $v_{zon}=\frac{\partial}{\partial
  x}<\phi>_{y}$ and $v_{dia}$, can produce a poloidal rotation
of the island. To investigate the role of these flows, we have plotted in Fig.~\ref{cap:figure7} the poloidal position of the center of the island $y_ {island}$,
and the poloidal positions related to the contributions of the diamagnetic velocity  
$v_{dia} $ and the zonal flow velocity $v_{zon}$. More precisely, we have plotted
$y_{dia}(t)=(1/\delta)\int_{0}^{t}dt\;\int_{-\delta/2}^{\delta/2}dx\;
v_{dia}(x,t)$ and
$y_{p,\phi}=(1/\delta)\int_{0}^{t}dt\;\int_{-\delta/2}^{\delta/2}dx\;(v_{dia}-v_{zon})$.
We observe that the model  $y_{dia} $ reproduces well the time evolution of the rotation of the island, before and after the bifurcation. {
At larger times, $t\gtrsim 6000\tau_{A}$ , we observe that the contribution of the zonal flow  cannot be neglected, even if the rotation of the island is mainly governed by the nonlinear generation of poloidal diamagnetic velocity.} In \cite{Ottaviani04}, a similar approach was taken, without an averaging over the sheet, and a value of the velocity at the center of the sheath was used. 

To summarize, we have shown that the dynamics of magnetic islands can be strongly affected by the presence of a background of interchange modes.  In the low resistivity and/or small $\Delta'$ limit, the coupling between the magnetic flux and the pressure is dominant compared to that between the magnetic flux and the plasma potential. In the asymptotic nonlinear regime, a pressure island structure builds up and the pressure pattern is not a flux function except in the center of the magnetic island.  In fact this regime is a result of a novel nonlinear transition that is observed for a wide range of parameters  when the condition $\eta<\eta_c$ is satisfied.  It is initiated by electrostatic interchange modes which compress the magnetic structure and generate small scales inside the island. Another noteworthy finding is that the bifurcation leads to a  change of symmetry of the diamagnetic velocity that occurs when the energy of the large scale interchange mode is of the same order of magnitude as the thermal energy contained in the cell maintaining the magnetic structure. The destabilization leads to a poloidal rotation of the island that is linked to the nonlinearly generated diamagnetic velocity in the current sheet. 
The basic phenomena highlighted by our results are reproducible over a large region of parametric  space and in that sense appear to be generic albeit within the constraints of our minimalist model. Effects ignored in our model including parallel heat conduction, parallel ion dynamics and contributions of the Hall and electron inertia terms may bring about some modifications. Investigation of such effects in an enlarged model are therefore necessary to provide a more global perspective of this complex phenomenon and for which our present studies provide a minimalist and basic description.

%
%













\end{document}